\documentclass[
 reprint,
 superscriptaddress,
 amsmath,amssymb,
 aps,
 prc,
 longbibliography,
 floatfix, 
 nofootinbib
]{revtex4-2}

\usepackage{amssymb}
\usepackage{amsmath}
\usepackage{MnSymbol}

\usepackage[colorlinks=true,
            linkcolor=blue,
            citecolor=blue,
            urlcolor=blue]{hyperref}

\usepackage{graphicx}
\usepackage{dcolumn}
\usepackage{bm}
\usepackage{url}
\usepackage{xcolor}
\usepackage{braket}
\usepackage{tikz}
\usepackage{comment}

\usetikzlibrary{positioning,arrows.meta,calc}

\begin{document}

\title{Optimizing artificial neural networks for dipole strength predictions in light nuclei}

\author{T.~Egert}
\email{tiegert@students.uni-mainz.de}
\affiliation{Institut f\"ur Kernphysik and PRISMA$^+$ Cluster of Excellence, Johannes Gutenberg-Universit\"at Mainz, 55128 Mainz, Germany}

\author{W.~G.~Jiang}
\email{wjiang@uni-mainz.de}
\affiliation{Institut f\"ur Kernphysik and PRISMA$^+$ Cluster of Excellence, Johannes Gutenberg-Universit\"at Mainz, 55128 Mainz, Germany}

\author{S.~Bacca}
\email{s.bacca@uni-mainz.de}
\affiliation{Institut f\"ur Kernphysik and PRISMA$^+$ Cluster of Excellence, Johannes Gutenberg-Universit\"at Mainz, 55128 Mainz, Germany}
\affiliation{Helmholtz-Institut Mainz, Johannes Gutenberg-Universit\"at Mainz, 55099 Mainz, Germany}

\date{\today}

\begin{abstract}
We present an optimized artificial neural network approach for predicting electric dipole strength functions in nuclei with $A < 50$. Building upon a previous global study [Phys.Rev.C {\bf 111} (2025) 5, L051308],  we focus here on the region of light nuclei where dipole responses are more structured. The new network incorporates a two-stage training process, a learned embedding of the proton number, explicit low-energy dipole onsets, uncertainty-weighted training, and high-energy regularization. Ensemble predictions show improved stability and substantially reduced variability across independently initialized networks compared with the earlier global neural network. Tests on selected isotopes withheld from training show that, for elements represented in the training set, the optimized network captures the main isotope dependent dipole strength systematics. As a further test, we compute electric dipole polarizabilities for selected light nuclei and compare them with literature values  revealing  a pronounced sensitivity to the covered energy interval. The resulting set of continuous electric dipole strength functions for nuclei with $A < 50$ provides a practical complement to existing tabulated photonuclear databases and is particularly suited for applications requiring smooth response functions over broad energy intervals. As an application, we provide an update on the electric dipole polarizability of $^9$Be.
\end{abstract}

\maketitle

\section{Introduction}
\label{introduction}
Artificial neural networks (ANNs) have become increasingly prominent across nuclear physics, with applications ranging from nuclear theory and experimental analysis to nuclear data science, as summarized in a recent review~\citep{Boehnlein:2021eym}. 
They have been used in nuclear structure and reaction studies to improve global mass-model systematics, emulate energy-density-functional calculations, and model inclusive electron-scattering response functions~\citep{Utama2016BNNMass,Niu2018BNNMass,Lasseri2020NuclearComplexity,Sobczyk2024}. ANNs have also been applied in ab initio nuclear theory to model-space extrapolations of  ground-state energies and radii in light nuclei~\citep{Jiang2019ANNExtrapolation, Knoll2025BenchmarkingANN}. These examples show that ANNs can learn correlated nuclear trends while benefiting from physics motivated inputs. 

In this work, we investigate ANN applications to the electric dipole response of nuclei, a key observable for probing nuclear dynamics across a wide range of excitation energies. The dipole response provides access to both collective and non-collective modes of motion~\cite{Bacca:2026tma} and enters energy-weighted sum rules that condense the strength distribution into physically meaningful quantities~\cite{LippariniStringari1989,BohigasLaneMartorell1979}. In particular, the inverse-energy-weighted sum rule defines the electric dipole polarizability $\alpha_D$, an observable that is especially sensitive to low-energy strength and provides important constraints on nuclear matter properties, including the symmetry energy and neutron-skin thickness~\citep{NEOS, NeutronSkin, NeutronStar, SymmetzryEnergy}.

This work builds on Ref.~\citep{OurPaper}, where an ANN was trained on electric dipole strength data across the nuclear chart. That global study demonstrated that neural networks can reproduce known dipole strength functions, expose tensions among datasets, and provide predictions for nuclei with incomplete experimental information. It also showed that the predicted responses can be used to extract integral observables connected to nuclear matter properties. The present paper follows the same general strategy, but addresses a different question: rather than maximizing coverage over the full nuclear chart, we ask how accurately the network can be optimized for light nuclei. This region poses a more stringent challenge because dipole responses are often more structured and because relatively small changes in the low-energy strength can have a visible impact on inverse-energy-weighted observables, such as the electric dipole polarizability. 

The motivation for this focus comes from precision studies of light nuclei, such as the charge-radius extractions in muonic atoms~\citep{Ohayon2024}. The interpretation of these measurements  requires nuclear-structure corrections, which depend sensitively on the electric dipole response function~\citep{JiReview}. A method that provides continuous dipole strength functions together with an estimate of ensemble variability for light nuclei is therefore particularly useful for these applications. By tailoring the network to this mass region, the present work aims to reduce the uncertainties that can arise when a model is optimized primarily for global coverage across the nuclear chart~\citep{OurPaper}.

Starting from the idea developed in~\cite{OurPaper}, in this work we restrict the training domain to nuclei with $A < 50$ and adapt the network to this mass region. The optimization includes an embedding of the proton number, residual connections, explicit threshold information, uncertainty weighted two-stage training, and a high energy regularization of the response. We use the optimized ANN to provide a continuous representation of the dipole response in light nuclei, together with an estimate of ensemble variability, and assess its performance against experimental data on dipole strength functions and polarizabilities for selected nuclei. The electric dipole polarizabiliy $\alpha_D$ is especially sensitive to low energy strength and has been measured or calculated with high precision in selected nuclei, see e.g.~Refs.~\citep{Birkhan2017AlphaDCa48, Miorelli2018AlphaDCa48}. It is therefore a useful test of whether the optimized response can be propagated to integral observables without relying on a direct interpolation of incomplete experimental energy windows. In this way, $\alpha_D$ serves as a representative application of the present framework, whose broader purpose is to provide stable ANN based dipole responses for light nuclei for later sum rule studies and for comparisons with experiment and theoretical calculations.

The paper is structured as follows. In Section~\ref{sec::structure}, we present the strategy we used to optimize the ANN to reproduce dipole response functions in light nuclei. In Section~\ref{results}, we present our results, and in Section~\ref{conclusions} we draw our conclusions.

\section{Problem statement and strategy}\label{sec::structure}
We train  ANNs exclusively on dipole-strength data to predict the electric dipole response $S_{D_1}$ for a given nucleus and quantify the associated ensemble variability. $S_{D_1}$ is defined as

\begin{align}
    S_{D_1}(\omega)= \frac{1}{2J_0 + 1} \sumint_{N \neq N_0, J}| \langle N_0 J_0 ||\hat{D}_1|| NJ \rangle | ^2\delta\, (\omega - \omega_N), \label{eq::SD1}
\end{align}

\noindent where $\sumint$ indicates a sum (integral) over discrete (continuous) nuclear excited states, $\omega_N$ is the excitation energy between the excited states $\ket{NJ}$ and the ground state $\ket{N_0J_0}$, and ${\langle N_0J_0 ||\hat{D}_1|| NJ \rangle}$ is the reduced matrix element of the electric dipole operator. The latter is defined as

\begin{equation}
    \hat{D}_{1}=\frac{1}{Z} \sum_{a=1}^{Z} R_a Y_{1}\!\left(\hat{\mathbf{R}}_a\right), \label{eq::dipoleOperator}
\end{equation}

\noindent where the sum runs over the $Z$ protons and the $\mathbf{R}_a$ are particle coordinates relative to the center of mass. The factor $1/Z$ follows the convention of Ref.~\cite{JiReview}. $S_{D_1}$ can be converted into the total photoabsorption cross-section $\sigma_\gamma(\omega)$ within the unretarded dipole approximation~\cite{BaccaPastore2014LightNuclei} using\footnote{The factor $Z^2$ is a consequence of the normalization in Eq.~\eqref{eq::dipoleOperator}.}

\begin{align}
    \label{sigma_gamma}
    \sigma_\gamma(\omega)= \frac{16\pi^3}{9}\alpha\,\omega\,Z^2 S_{D_1}(\omega).
\end{align}

\noindent The  electric dipole polarizability is an inverse energy-weighted sum rule of the dipole strength function which can be obtained from the photoabsoprtion cross section as 

\begin{align}
    \label{alphaD}
    \alpha_D = \frac{\hbar c}{2 \pi^2}\int_0^{\infty} d\omega \, \frac{\sigma_{\gamma}(\omega)}{\omega^2}\,.
\end{align}

While a plain feedforward ANN was employed in our previous work~\citep{OurPaper}, the present study adapts and optimizes the network basis specifically for light nuclei. We train on an evaluated photoabsorption compilation from the IAEA Photon Strength Function database\footnote{\url{https://nds.iaea.org/PSFdatabase-legacy/}}, described in~\citep{IAEADatabase, IAEA2, IAEA3, IAEA4}. The backbone comprises data for 152 nuclei with $Z=2$–$94$ across 25 elements, but, since this work focuses on light nuclei, the dataset is restricted to $A < 50$. The network is therefore not trained to learn effects that are specific to heavy nuclei. To better cover the low-energy region, we add the Oslo–method data~\citep{Oslo} and NRF measurements~\citep{NRF}, as done as in~\citep{OurPaper}. Targeted updates are included for selected isotopes (e.g. $^{40,48}\mathrm{Ca}$) with the addition of independent Coulomb-excitation reactions in forward angle $(p,p')$~\citep{PPprime}. After compiling the experimental information, each dataset and its associated uncertainties are interpolated separately within its measured energy interval using shape-preserving piecewise cubic Hermite interpolation. The number of interpolation points is proportional to the covered energy range, increasing the training sample without extrapolating beyond the measured intervals. With respect to our earlier work~\cite{OurPaper}, several improvements have been implemented, which are outlined in the following.

Understanding the implemented method requires connecting the physical structure of the dipole response with the computational design of the ANN. From a physics perspective, the response is considered in three energy regions. These comprise the low-energy region, the giant dipole resonance (GDR), and the high-energy region. Across the compiled dataset, the experimental coverage, defined as the range and density of photon energies for which measurements are available, is strongly concentrated around the GDR~\citep{IAEADatabase,IAEA2,IAEA3,IAEA4}. NRF and Oslo-method measurements provide additional low-energy information for selected nuclei~\citep{NRF,Oslo}. Nevertheless, the response of many nuclei remains only weakly constrained below the GDR and at high excitation energies. This presents a particular challenge because the behavior of an ANN outside the domain represented by its training data is not generally determined by the data and may become unreliable without additional constraints~\citep{Jiang2019ANNExtrapolation,Pastore:2020uso,Boehnlein:2021eym}. From a computational perspective, two aspects must therefore be considered. The network architecture determines how the physical inputs are represented and transformed into a prediction. The training procedure determines how the network parameters are learned from the available data through the construction of training samples, the loss function, the weighting scheme, and the optimization and validation procedures. The ANN architecture and the training procedure were developed together through a close examination of the experimental dataset and iterative diagnostic studies.

Before specifying how the three energy regions are treated during training, a suitable network architecture must first be established. The architecture provides a common functional representation of the dipole response over the full energy range. It does not explicitly divide the response into low-energy, GDR, and high-energy regions or prescribe different behavior in them. Instead, it influences the result indirectly through its representational capacity, regularization, information flow, and output constraints. It therefore forms the foundation on which the region-dependent treatment is built. The available data and the additional physical constraints are subsequently introduced through the construction and weighting of the training objective.

\begin{figure}[h]
  \centering
  \includegraphics[width=0.8\linewidth]{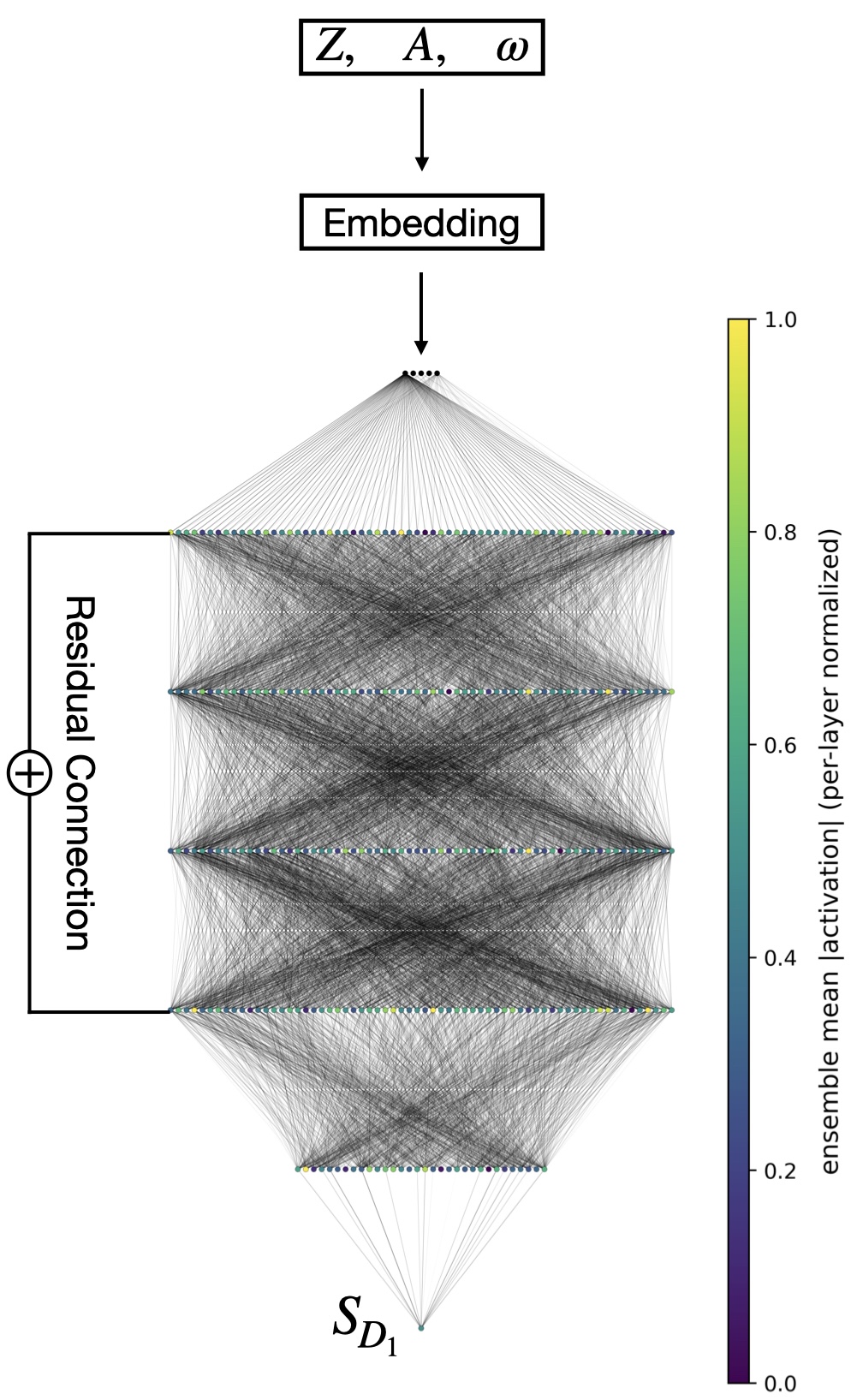}
  \caption{Schematic visualization of the ANN at hand. The line thickness of the connections between the layers corresponds to the absolute training weight of the respective nodes.}
  \label{fig:NN_structure}
\end{figure}

The network architecture comprises five fully connected hidden layers. As shown in  Fig.~\ref{fig:NN_structure}, The first four layers contain 64 neurons each and are followed by a 32-neuron bottleneck layer, introduced to suppress inactive neurons identified in diagnostic studies. The input feature set consists of $A$, $Z$, and $\omega$, where $A$ is the mass number, $Z$ is the proton number, and $\omega$ is the photon energy. The proton number is represented by a trainable embedding layer that maps every integer value of $Z$ onto a three-dimensional latent vector shared by all isotopes of the corresponding element (Supplemental Material Sec.~S1 A). Intuitively, this assigns each element a compact set of numerical descriptors that are learned during training, rather than treating $Z$ only as an ordinary scalar input~\citep{EmbeddingInPhysics}. Details about the embedding method and its implementation as well as a visualization of the corresponding embedding space can be found in the supplementary material~\citep{SupplementalMaterial}. The mass number is retained as a scalar input and converted to floating-point representation before being concatenated with the embedded proton number and the photon energy. The resulting five-dimensional feature vector is standardized by a batch-normalization layer before entering the first fully connected layer. Each of the four 64-neuron layers consists of a linear transformation followed by batch normalization and an exponential linear unit (ELU) activation. The ELU activations improve gradient flow and training stability in the presence of sparse and heterogeneous nuclear data. A residual connection adds the activated output of the first 64-neuron layer to the output of the fourth 64-neuron layer before the combined representation is passed to the 32-neuron bottleneck. The bottleneck also uses an ELU activation. An $L_2$ kernel regularizer, implemented using Keras~\citep{KerasDoc}, is applied to all four 64-neuron layers and to the bottleneck layer. It adds a penalty proportional to the squared Euclidean norm of the kernel weights, thereby discouraging excessively large parameter values~\citep{L2reg}. The ANN output is produced by a one-dimensional positive-valued head using a softplus activation to enforce physical non-negativity of the dipole strength. The selection of the input features is crucial for network performance. As done in Ref.~\citep{OurPaper}, we use a minimal input configuration consisting only of $(A,Z,\omega)$. This input configuration allows users to obtain predictions for arbitrary isotopes of elements represented during training by specifying only the mass number, proton number, and desired energy grid.

Model-to-model variability is assessed using an ensemble of 100 independently initialized networks trained with stochastic data shuffling~\citep{ensembleLearning1,ensembleLearning2,ensembleLearning3,ensembleLearning4}. The final prediction is given by the ensemble median, while the reported highest-density intervals quantify the variability among the accepted networks under the fixed modelling assumptions of this work. Integrated quantities in the tables are reported with $68\%$ high density intervals (HDIs) to facilitate comparison with commonly quoted one-standard-deviation experimental uncertainties, whereas the response-function figures show $90\%$ HDIs to make the energy dependence of the ensemble variability more visible.

To assess isotope-level transfer, all response data for $^{14}\mathrm{C}$, $^{15}\mathrm{N}$, $^{17}\mathrm{O}$, and $^{25}\mathrm{Mg}$ were withheld from training, validation, and model-acceptance procedures and reserved for the subsequent evaluation of predictions for unseen isotopes of elements represented during training.

With this architectural foundation established, we now consider the three physical energy regions and the ANN training methods used to obtain stable predictions in each. Although the following discussion is organized according to three physical energy regions, these regions are not optimized independently. A single ANN represents the dipole response over the full energy range, and all loss contributions act on the same network parameters. During the first training stage, the ANN is fitted to the complete supervised dataset using an uncertainty-weighted data term together with additional terms for peak preservation and subthreshold suppression. The second stage retains these contributions and introduces a preservation term and additional constraints on the high-energy tail. The regional organization therefore identifies where each component has its primary effect rather than dividing the optimization into three separate problems. Detailed definitions of the loss functions and the two-stage training procedure are provided in the Supplemental Material~\citep{SupplementalMaterial}.

\subsection{Low energy region}\label{sec::LowEnergies} 
As already mentioned,  experimental strength data are concentrated predominantly around the GDR and, with a few exceptions, do not cover the complete energy interval down to the lowest possible $\omega$ or the asymptotic high-energy region. The missing regions nevertheless contribute to sum rules obtained by integrating the full response, with the low-energy strength being particularly important for inverse-energy-weighted observables such as  $\alpha_D$. 

To regularize the ANN in the low energy region, we define a nucleus dependent low-energy onset $\omega_{\mathrm{th}}$. This energy is taken as the particle emission threshold unless a known (discrete) dipole transition occurs at a lower excitation energy, in which case the energy of that transition is used. Synthetic data with $S_{D_1}=0$ are introduced only below $\omega_{\mathrm{th}}$. This construction avoids imposing a vanishing response across known bound dipole transitions below the first open particle channel. Experimental Oslo method and NRF strengths below the threshold are retained in the dataset whenever available. The number of synthetic points is balanced among the different nuclei to prevent nuclei with threshold information from being overrepresented during training. Each synthetic point is also assigned a finite uncertainty of $10$\% of median uncertainty of the respective nucleus, thereby avoiding formally infinite sample weights in the inverse-uncertainty weighting scheme. The suppression of the response below the first known dipole transition is further reinforced through an additional one-sided loss term that penalizes any positive ANN prediction at the synthetic subthreshold points.

\subsection{GDR region}
The GDR region contains the largest concentration of experimental data and therefore provides the primary anchor for the predicted response. In contrast to the weakly constrained low- and high-energy regions, no prescribed functional form is imposed on the response in this region. Instead, the ANN learns the resonance structure from the database through the uncertainty-weighted data term. This weighting is particularly relevant when measurements from different experiments overlap but carry different uncertainties (see Supplemental Material~\cite{SupplementalMaterial}, Sec.~S1 B).

The pointwise data loss is supplemented by an asymmetric, one-sided peak penalty, because points near the peak constitute only a small fraction of the training sample compared with the more densely sampled surrounding response. Consequently, minimizing the ordinary squared error alone can favor a smoother prediction that systematically underestimates the resonance maximum. The additional term penalizes underprediction at selected high-strength points, while the symmetric data term continues to also penalize overprediction. The selection is based on the strength relative to a robust nucleus-dependent reference value within the resonance energy interval and does not impose a particular peak shape or position (see Supplemental Material~\cite{SupplementalMaterial}, Sec.~S1 C). A corresponding acceptance criterion verifies that the dominant resonance strength is retained after each training stage. During the second stage, the preservation term additionally limits changes to the accepted supervised response while the high-energy behavior is regularized (see Supplemental Material~\cite{SupplementalMaterial}, Sec.~S1 D).

\subsection{High energy region}
At high excitation energies, the experimental coverage becomes increasingly sparse and the continuation of the response is no longer sufficiently constrained by supervised data. Within the quasi-deuteron (QD) picture, the photoabsorption cross section is written as $\sigma_\gamma(\omega)=\sigma_{\mathrm{GDR}}(\omega)+\sigma_{\mathrm{QD}}(\omega)$, where the high-energy contribution arises predominantly from correlated neutron-proton pairs~\citep{LevingerQuasiDeuteron,HighEnergyPhotoEffectLevinger,IAEA2019Photonuclear}. The corresponding dipole strength is expected to approach an asymptotic dependence proportional to $\omega^{-5/2}$~\citep{IAEA2019Photonuclear}. We use this behavior to constrain the shape of the extrapolated response without prescribing its absolute normalization or imposing a common transition energy for all nuclei.

The high-energy constraint is introduced during the second training stage, after the ANN has learned the measured response. For each nucleus, the accepted first-stage prediction is used to identify the energy above which the high-energy tail begins. The resulting onset energies are then smoothed across the training nuclei to obtain a consistent dependence on nuclear properties. Geometrically spaced pairs of collocation energies are then placed beyond the highest positive-strength energy represented by the supervised data. These points are not assigned training data. Instead, the loss constrains the ratio of the predictions at neighboring energies to follow the expected $\omega^{-5/2}$ dependence and penalizes local increases in the tail. The constraint is activated smoothly across the transition to the asymptotic region. During this fine-tuning stage, the supervised loss remains active and a preservation term limits changes to the accepted first-stage response. Each final network is additionally required to produce a finite and smoothly decreasing high-energy response consistent with the expected QD behavior (see Supplemental Material~\cite{SupplementalMaterial}, Secs.~S1 C and S1 D).

\section{Results}\label{results}
\begin{figure}[tbp]
  \centering
\includegraphics[width=\linewidth]{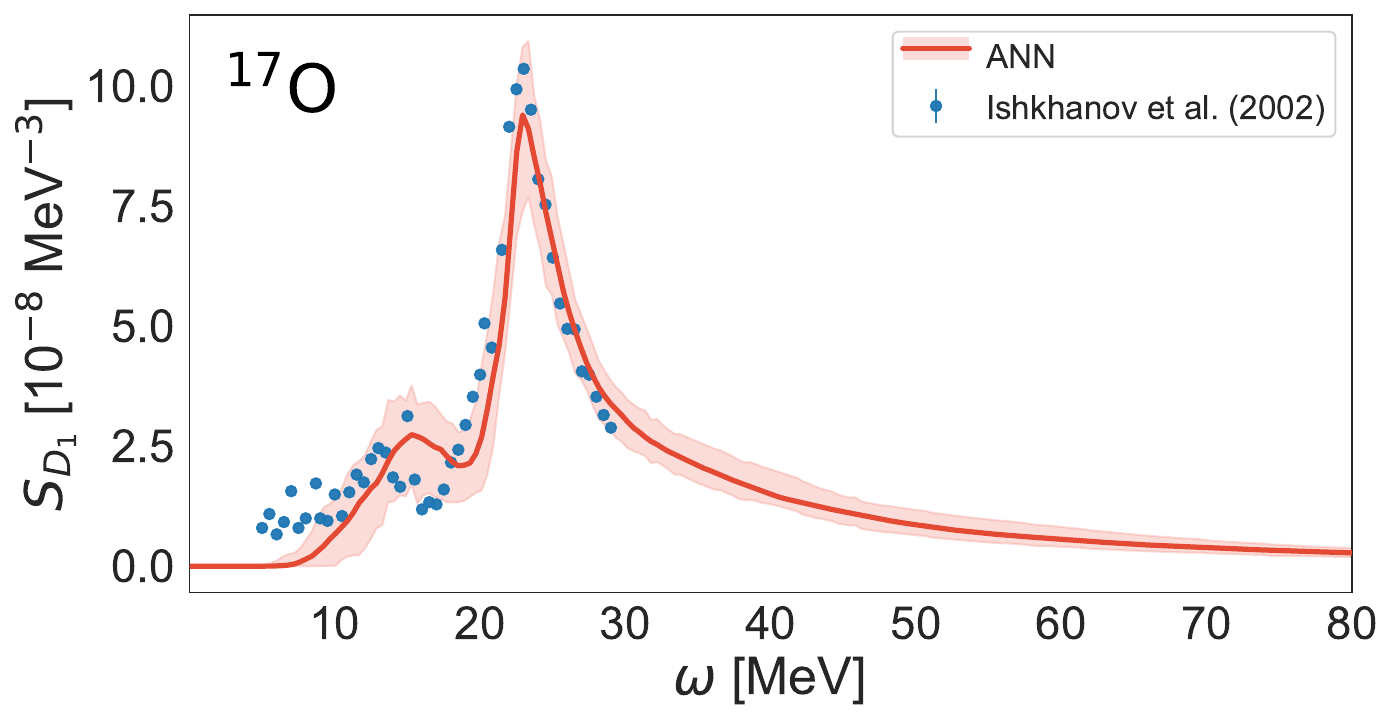}
  \vspace{0.5em} 
\includegraphics[width=\linewidth]{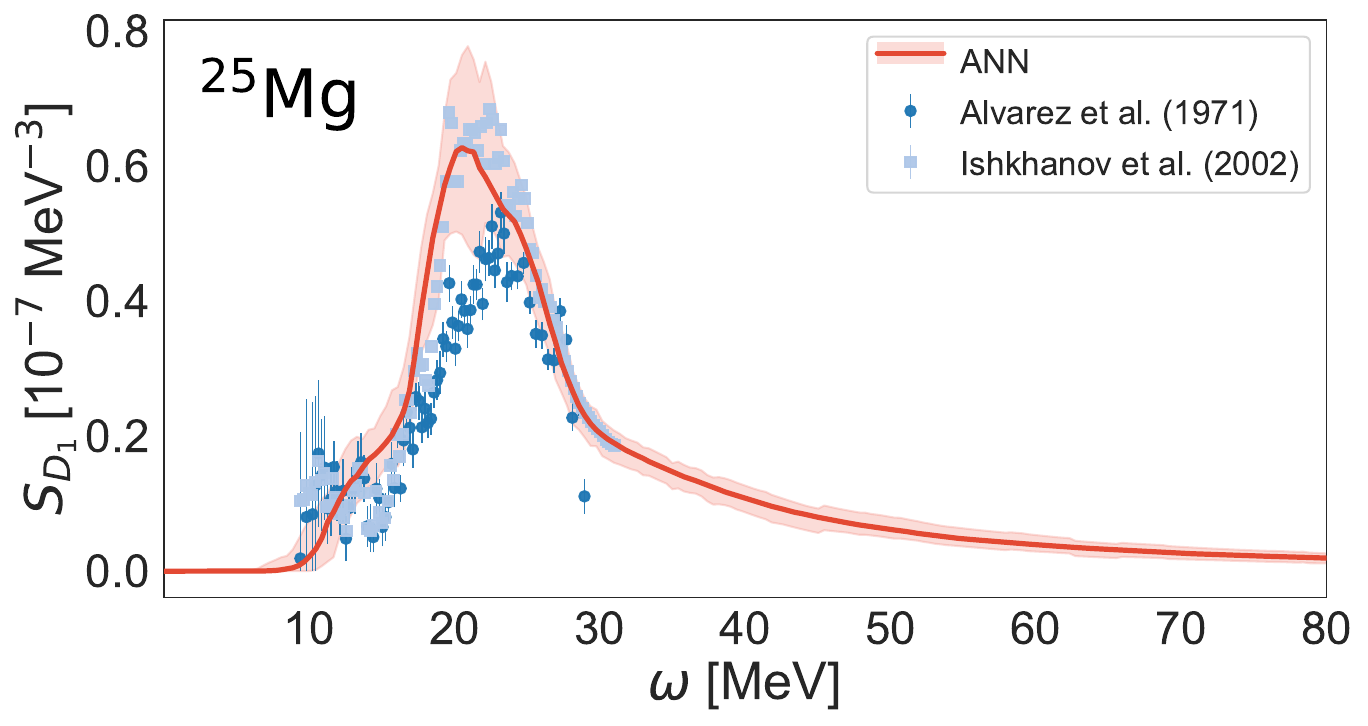}
  \caption{ANN predictions for $^{17}$O  (upper panel) and $^{25}$Mg (lower panel).  Experimental data are from~\citep{Ish2002, 1971Alv}.}
  \label{fig:testnuclei}
\end{figure}

We start the presentation of our results by discussing the ANN predictions for test nuclei. As mentioned above, $^{14}\mathrm{C}$, $^{15}\mathrm{N}$, $^{17}\mathrm{O}$, and $^{25}\mathrm{Mg}$ were excluded from the training data. To evaluate the transfer of the learned response to other nuclei, in Figure~\ref{fig:testnuclei} we compare the ANN predictions for $^{17}\mathrm{O}$ and $^{25}\mathrm{Mg}$ against available experimental data, not seen in the training phase. 

\noindent As one can see, for $^{17}\mathrm{O}$ the ANN captures the double peak structure and predicts both the peak positions and their amplitudes well. For $^{25}\mathrm{Mg}$, the ANN also reproduces the main peak position and amplitude. In this case the prediction appears to favor the Ishkhanov~\citep{Ish2002} dataset over the Alvarez~\citep{1971Alv} dataset. Since $^{25}\mathrm{Mg}$ data were excluded from training, this preference cannot be caused by the small uncertainties of the Ishkhanov data. It must rather arise from the systematics learned from neighboring nuclei and from the global constraints imposed by the training procedure.

It is important to remark that, even if  $^{14}\mathrm{C}$, $^{15}\mathrm{N}$, $^{17}\mathrm{O}$, and $^{25}\mathrm{Mg}$ were excluded from training, the corresponding element remains represented by at least one other isotope. The test therefore probes isotope-level transfer within isotope-chains known to the network, rather than generalization to previously unseen elements. This setup is particularly informative in the presence of element-wise embedding (see Supplemental Material~\citep{SupplementalMaterial}), since it directly tests how well the learned element representation transfers to unseen isotopes of the same element. These results support the ability of the network to interpolate between neighboring isotopes and to perform  predictions along represented isotope chains. 

With the completion of the testing phase, the previously excluded nuclei are reintroduced into the training dataset. All networks are then retrained using the full dataset in order to maximize predictive accuracy for the final dipole strength predictions. This final training step ensures that the ANN benefits from the complete available dipole strength information when producing the results used for the sum rule analysis.

\begin{figure}[h]
  \centering
  \includegraphics[width=\linewidth]{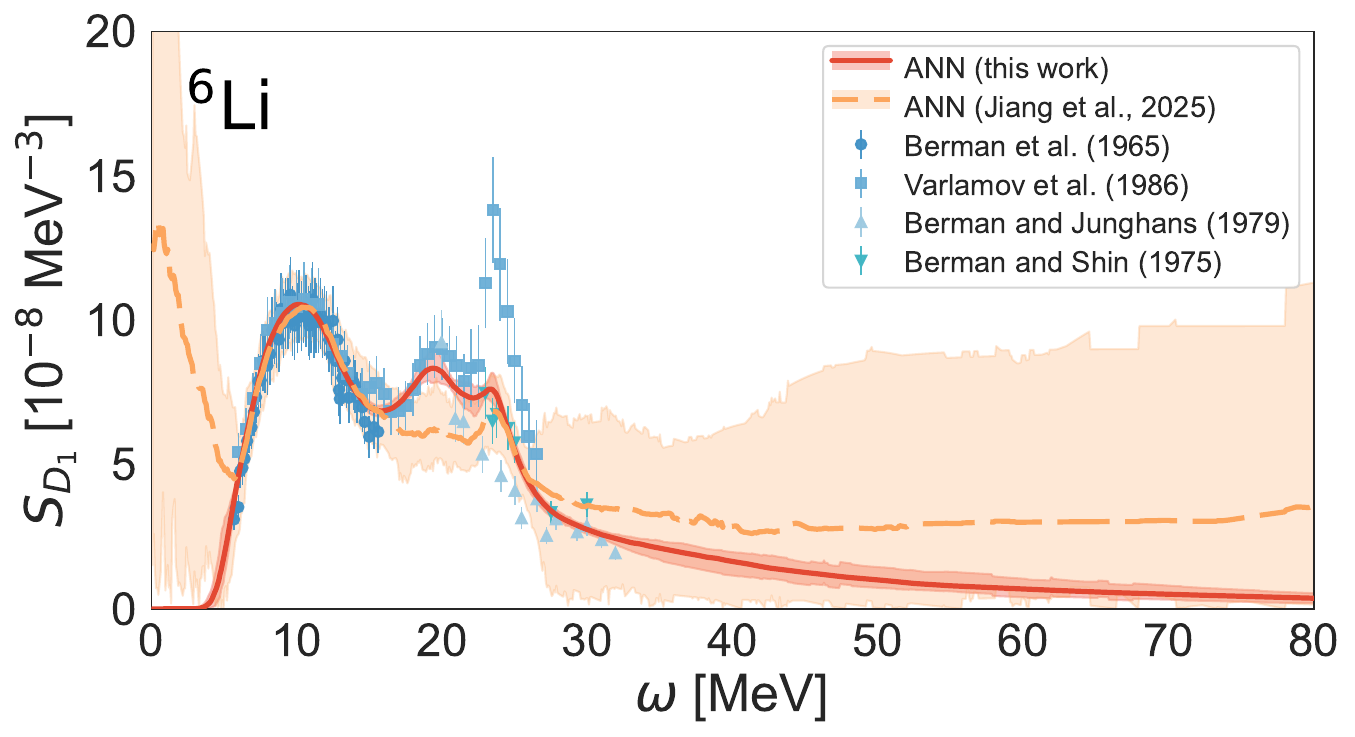}
  \caption{Comparison of the predictive performance of this works's ANN and that from Jiang et al.~\citep{OurPaper} for $^6$Li. Experimental data are taken from Refs.~\citep{1965Be1, 1986Var, Junghans1979, Shin1975}.}
  \label{fig:oldnew}
\end{figure}

At this point it is interesting to compare the effect that the ANN optimization on light nuclei has in comparison to the global ANN introduced in Ref.~\cite{OurPaper}. Figure~\ref{fig:oldnew} shows such comparison for $^6$Li, as a representative examples of a light nucleus.  For both ANNs, the shaded band show the point-wise $90\%$ HDI of the predictions from 100 accepted ensemble members. Clearly, the combined modifications introduced in this work yield predictions with substantially smaller ensemble variability than the earlier ANN.  The improvement is most visible close to threshold, where the new ANN remains consistent with the physically motivated constraint outlined in Sec.~\ref{sec::structure} A of a vanishing response below threshold and develops a much narrower ensemble variability band in the first strength region. This improved behavior is going to affect inverse energy-weighted sum rules such as the electric dipole polarizability. Between about $10~\mathrm{MeV}$ and $20~\mathrm{MeV}$, the new ANN follows the main rise and fall of the measured response while showing less spread between ensemble members than the ANN of Ref.~\cite{OurPaper}. In the region from about $20~\mathrm{MeV}$ to $30~\mathrm{MeV}$, both ANNs do not reproduce the additional structure suggested by the Varlamov dataset. This is consistent with the tension between that dataset and the more precise Berman's measurements in the same energy interval, as well as with the constraints learned from neighboring nuclei. Above about $30~\mathrm{MeV}$, the ANN of Ref.~\cite{OurPaper} develops a much broader tail uncertainty, whereas the newly-optimized ANN gives a smoother and more controlled falloff at high excitation energy, thanks to the implemented second stage custom loss, where the QD model has been imposed as outlined in Sec.~\ref{sec::structure} C. The reduced ensemble variability in the high-energy tail is particularly important for sum-rule studies involving energy weights that enhance the contribution of the high-energy region.

\begin{table}[h]
\centering
\begin{tabular}{llll}
\hline
Nucl. & Ref.~\citep{OurPaper} &  This work & Experiment\\
\hline
$^{7}\mathrm{Li}$  & $1.1^{+0.2}_{-0.5}$ & $0.215^{+0.003}_{-0.003}$ & $0.196(2)$~\citep{AHRENS} \\
$^{9}\mathrm{Be}$  & $1.4^{+0.2}_{-0.6}$ & $0.454^{+0.013}_{-0.008}$  & $0.192(5)$~\citep{AHRENS} \\
$^{12}\mathrm{C}$  & $1.5^{+0.4}_{-0.8}$ & $0.302^{+0.010}_{-0.017}$  & $0.313(5)$~\citep{AHRENS}\\
$^{16}\mathrm{O}$  & $1.5^{+0.7}_{-0.6}$ & $0.418^{+0.020}_{-0.010}$  & $0.580(9)$~\citep{AHRENS} \\
$^{40}\mathrm{Ca}$ & $3.0^{+0.5}_{-0.8}$ & $1.76^{+0.01}_{-0.01}$     & $2.22(3)$~\citep{AHRENS} \\
 &  &      & $1.87(3)$~\citep{Birkhan2017AlphaDCa48}\\
  &  &      & $1.92(17)$ ~\citep{Fearick:2023lyz}\\
 $^{48}\mathrm{Ca}$ & $3.6^{+0.5}_{-1.0}$ &  $2.43^{+0.02}_{-0.05}$ &  $2.07(22)$~\citep{Birkhan2017AlphaDCa48}\\
\hline\hline
 &  & & Empirical~\citep{empiricalalphaD2} \\
 \hline
$^{17}\mathrm{O}$  & $1.6^{+0.9}_{-0.6}$ & $0.565^{+0.010}_{-0.012}$  & $0.622$ \\
$^{18}\mathrm{O}$  & $1.8^{+0.7}_{-1.1}$ & $0.569^{+0.016}_{-0.016}$  & $0.661$ \\
$^{19}\mathrm{F}$  & $1.5^{+0.8}_{-0.8}$ & $0.79^{+0.01}_{-0.01}$     & $0.700$ \\
$^{23}\mathrm{Na}$ & $1.7^{+1.0}_{-0.7}$ & $0.72^{+0.03}_{-0.03}$     & $0.870$ \\
$^{24}\mathrm{Mg}$ & $1.8^{+0.9}_{-0.8}$ & $0.80^{+0.04}_{-0.03}$     & $0.915$ \\
$^{25}\mathrm{Mg}$ & $1.8^{+0.9}_{-0.9}$ & $0.87^{+0.02}_{-0.02}$     & $0.961$ \\
$^{26}\mathrm{Mg}$ & $1.9^{+0.7}_{-1.0}$ & $1.11^{+0.03}_{-0.03}$     & $1.01$ \\
\hline
\end{tabular}
\caption{Electric dipole polarizabilities in fm$^3$: ANN predictions from Ref.~\cite{OurPaper} and this work compared to experimental data and empirical estimates.}
\label{tab:alphaD-exp-ann}
\end{table}

We now turn our attention to the  electric dipole polarizability. Table~\ref{tab:alphaD-exp-ann} compares $\alpha_D$ results from the ANN of Ref.~\citep{OurPaper} with the optimized ANN developed in this work. For both models, the response is obtained using Eq.~(\ref{alphaD}) integrating from $0$ to $100~\mathrm{MeV}$, and the quoted non-symmetric intervals correspond to the $68\%$ HDIs of the respective ensemble distributions. The optimized ANN exhibits substantially less variation across the ensemble members than the model of Ref.~\citep{OurPaper}. This improvement is also evident in the electric dipole polarizability $\alpha_D$, whose inverse-energy weighting makes it particularly sensitive to the low-energy dipole response. The reduced variability of $\alpha_D$ is therefore consistent with the more stable low-energy predictions illustrated for $^6$Li in Fig.~\ref{fig:oldnew}. We compare these numbers also to available experimental data from Ref.~\cite{AHRENS,empiricalalphaD2,Fearick:2023lyz}. Their statistical uncertainties are written in parentheses. For nuclei with mass number $16<A<40$, where no direct experimental data are available,
 we compare to the empirical estimate of a hydro-dynamical model from Refs.~\citep{empiricalalphaD2}(see also \citep{empiricalalphaD1,empiricalalphaD3,empiricalalphaD4}). Compared with the ANN of Ref.~\citep{OurPaper}, the predictions of the newly optimized ANN lie considerably closer to the available experimental measurements and empirical estimates. One can also notice that the difference of the Ref.~\cite{OurPaper} ANN results  with respect to the experimental or empirical values gets smaller as $A$ increases. This is expected, since that global ANN approach was optimized for broad coverage across the nuclear chart rather than specifically for light nuclei.

Overall, the ANN optimized for light nuclei reproduces the mass-number dependence reasonably well. Compared with experimental data, it shows both under- and overprediction, with the largest discrepancy occurring for $^9$Be, and the predictions do not always agree within the quoted uncertainties. The ANN also tends to slightly underestimate the hydrodynamic-model estimates. For $^{40}$Ca, three different experimental values are reported, not always compatible with each other. While the original experimental paper~\cite{AHRENS} quotes $\alpha_D=2.22(3)$ fm$^3$, as already noticed in Ref.~\cite{Bacca2014} integrating the original data one actually obtains a smaller value amounting to $\alpha_D=1.87(3)$ fm$^3$~\cite{Birkhan2017AlphaDCa48}.  Recent data measured with a $(p,p')$ experiment~\cite{Fearick:2023lyz} led to $\alpha_D=1.92(17)$ fm$^3$, which is actually in agreement with the new ANN prediction within uncertainties.
 
It is worth further investigating possible source of the observed differences with the experimental data. Upon inspecting Eq.~(\ref{alphaD}), an important aspect is what integration limits are actually implemented. Theoretically, one should integrate from zero to infinity, but experimental data do not cover an infinitely large energy-range. The $\alpha_D$ values from Ref.~\citep{AHRENS}, determined from a photo-absorption attenuation  experiment, were obtained by integrating between $10$ and $100~\mathrm{MeV}$. The ANN values in Table~\ref{tab:alphaD-exp-ann} instead extend down to $0~\mathrm{MeV}$. Most importantly, the ANN is trained on the IAEA database~\citep{IAEADatabase}, not exclusively on the original Ahrens et al.~\citep{AHRENS}. Before inclusion in the IAEA database, all cross sections were evaluated and restricted to nucleus-dependent energy intervals. Data points were omitted at low energies when the estimated $(\gamma,\gamma)$ contribution exceeded $10\%$ because the photo-particle cross sections used to extract the photon strength function do not include this channel and would consequently yield an incorrectly small strength. Furthermore, high-energy points were omitted when the estimated quasi-deuteron contribution exceeded $10\%$ to exclude the region in which many particle-emission channels are open. Consequently, the training set includes only a subset of the data from Ref.~\citep{AHRENS} that contributed to the published $\alpha_D$ values, together with additional measurements from other experiments.

In particular, the database \citep{IAEADatabase} excluded energies below $7.34~\mathrm{MeV}$ and energies above $27.10~\mathrm{MeV}$ for $^7$Li and the high-energy region above $28.50~\mathrm{MeV}$ for $^{9}\mathrm{Be}$. For $^{12}\mathrm{C}$, it excludes energies below $18.750~\mathrm{MeV}$ and above $27.45~\mathrm{MeV}$, while for $^{16}\mathrm{O}$ the corresponding boundaries are $15.85$ and $30.50~\mathrm{MeV}$. The natural-calcium entry, which is dominated by $^{40}\mathrm{Ca}$, documents a low-energy boundary of $15.74~\mathrm{MeV}$ and terminates at $30.82~\mathrm{MeV}$. 
It is therefore interesting to compute $\alpha_D$ using only this reduced energy-range in the integral of Eq.~(\ref{alphaD}).

\begin{table}[h]
\centering
\begin{tabular}{lccc}
\hline
\shortstack{Nucl.\\\strut} &
\shortstack{ Interval\\$[\mathrm{MeV}]$} &
\shortstack{$\alpha_D$ from ~\citep{AHRENS} \\$[\mathrm{fm}^3]$} &
\shortstack{$\alpha_D$ from ANN\\$[\mathrm{fm}^3]$} \\
\hline
$^{7}\mathrm{Li}$ & $7.34$--$27.10$ & $0.159$ & $0.159^{+0.001}_{-0.002}$ \\
$^{9}\mathrm{Be}$ & $10.50$--$28.50$ & $0.120$ & $0.126^{+0.002}_{-0.001}$ \\
$^{12}\mathrm{C}$ & $18.75$--$27.45$ & $0.182$ & $0.166^{+0.001}_{-0.001}$ \\
$^{16}\mathrm{O}$ & $15.85$--$30.50$ & $0.369$ & $0.301^{+0.002}_{-0.002}$ \\
$^{40}\mathrm{Ca}$ & $15.74$--$30.82$ & $1.43$ & $1.492^{+0.005}_{-0.009}$ \\
\hline
\end{tabular}
\caption{Electric dipole polarizabilities obtained by integrating \citep{AHRENS} data and the light-nuclei optimized ANN prediction over identical energy intervals. The ANN entries show the ensemble median and $68\%$ HDI.}
\label{tab:alphaD-ahrens-common-window}
\end{table}

Table~\ref{tab:alphaD-ahrens-common-window} shows $\alpha_D$ obtained from integrating the 
\citep{AHRENS} data and the light-nuclei optimized ANN over the same nucleus-dependent interval indicated in the second column, following the IAEA-adopted prescription. 
The entries in the third column are derived from a shape-preserving interpolation of the tabulated \citep{AHRENS} data. We report only the central value without propagating uncertainties, as we deem it sufficient for the present test. The last column is obtained by integrating each member of the new ANN ensemble, reporting the median together with the upper and lower limits of the $68\%$ HDI. 
 
 Using identical integration limits  provides a more direct comparison of their predicted response within the experimentally represented region and dramatically reduces the numerical differences, which are now well below 10$\%$ for all nuclei with the exception of $^{16}$O, where the relative difference is approaching 20$\%$. For $^{7}\mathrm{Li}$, the agreement over the common interval is nearly exact, while the reduction of relative difference going from Table~\ref{tab:alphaD-exp-ann} to Table~\ref{tab:alphaD-ahrens-common-window} is most pronounced for $^{9}\mathrm{Be}$. Overall, the comparison in Table~\ref{tab:alphaD-ahrens-common-window} shows that differences in energy coverage account for a substantial part of the discrepancies. It is important to notice that exact agreement with Ref.~\citep{AHRENS} is not expected even after matching the energy intervals. The ANN is not fitted to the data of Ref.~\citep{AHRENS} in isolation, but to all experimental datasets available in the same energy region using their assigned uncertainties. Where these measurements differ, the predicted ANN response represents an uncertainty-weighted compromise rather than a direct interpolation of Ref.~\citep{AHRENS}. The shared ANN representation also incorporates correlations learned from other nuclei. The remaining differences in Table~\ref{tab:alphaD-ahrens-common-window} are therefore consistent with differences among the experimental datasets and with the cross-nuclear systematics learned during training.

\begin{figure}[tbp]
  \centering
\includegraphics[width=\linewidth]{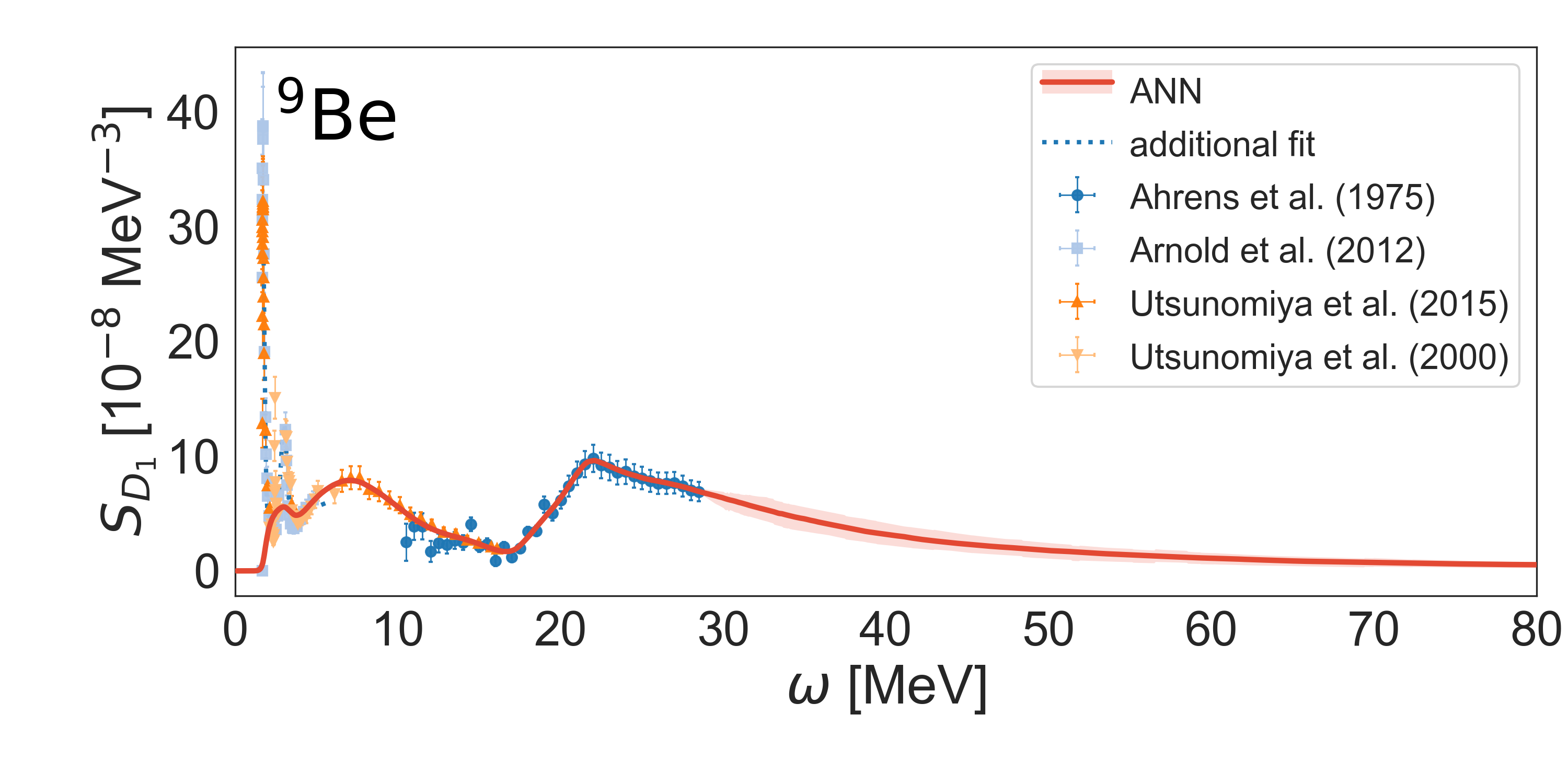}
\includegraphics[width=\linewidth]{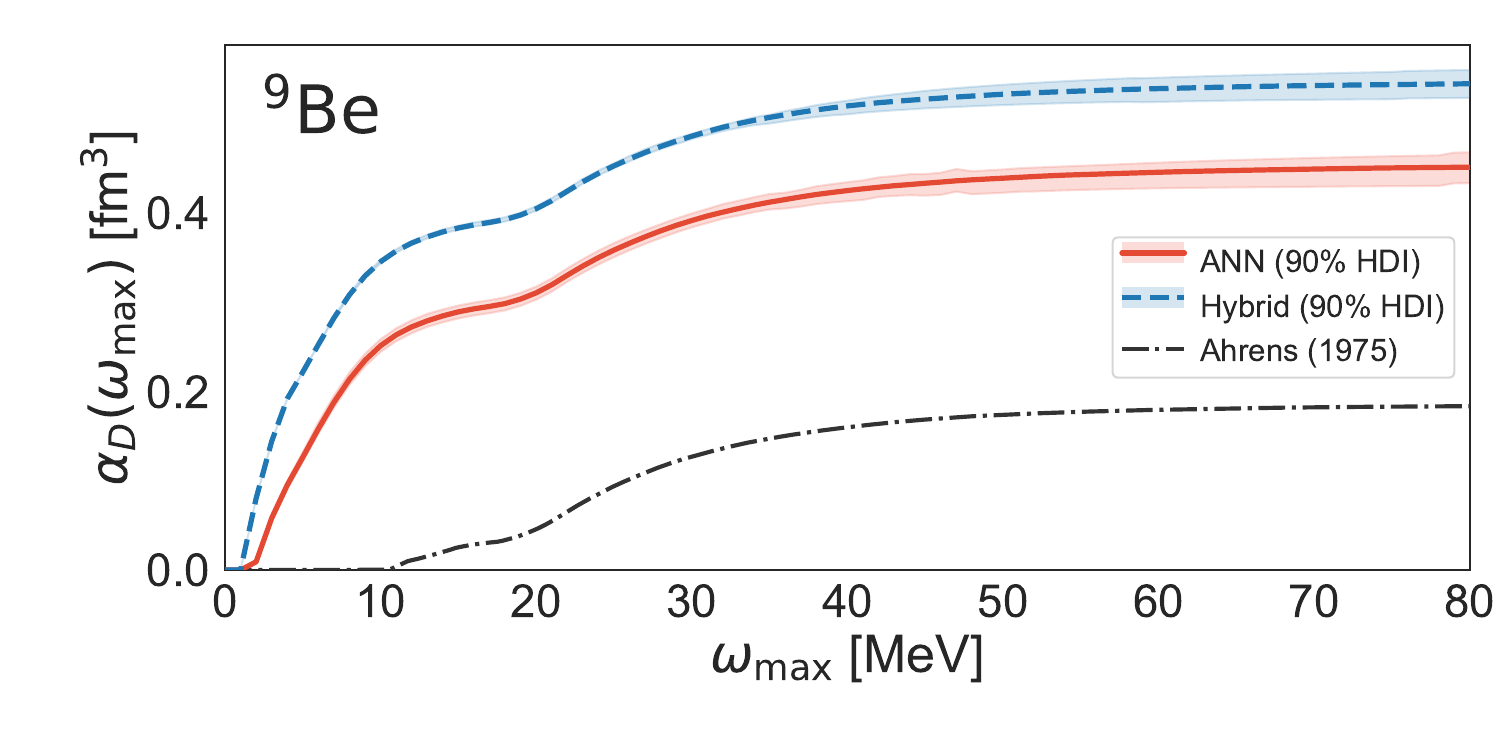}
  \caption{Dipole strength (upper panel) and running polarizability sum rule (bottom panel) for $^9\mathrm{Be}$. ANN predictions are shown along with a fit to the resonant ($1/2^+, 5/2^+$ and $3/2^+$) data and all of the available experimental data from Refs.~\citep{AHRENS, Arnold, Uts2000, Uts2015} (see text for details).}
  \label{fig:Be9}
\end{figure}

The  $^{9}\mathrm{Be}$ case warrants closer inspection. When integrated over the full energy range, the ANN value for $\alpha_D$ is more than twice larger than the value quoted by~\citep{AHRENS}, whereas the relative difference is reduced to $5\%$ when both responses are integrated over the common database interval.  This indicates that the difference originates predominantly from low-energy strength outside the shared energy interval.
Figure~\ref{fig:Be9} shows a comparison of the ANN prediction for the dipole response function to available experimental data. One can clearly see that from the experimental stand point there is a pronounced concentration of dipole strength below $10~\mathrm{MeV}$, i.e., outside of the range considered in Table~\ref{tab:alphaD-ahrens-common-window}. This strength contributes to the ANN integral over the full energy range but lies below the lower integration limit used for the literature value from Ref.~\citep{AHRENS}. Its contribution to $\alpha_D$ is further enhanced by the inverse energy weighting, as can be seen in the bottom panel of Figure~\ref{fig:Be9}, where we present the running $\alpha_D$ sum rule using Eq.~(\ref{alphaD}) and integrating up to a maximal energy value $\omega_{\rm max}$.

It is also evident that the ANN underestimates the height of the first two narrow peaks below 5 MeV. Resolving such a localized feature may require greater model flexibility or a training strategy that assigns it additional influence. Since $^9\mathrm{Be}$ is the only nucleus in the present dataset with such a pronounced narrow structure at low energy, this could reduce the ability of the model to describe common systematics across nuclei. The comparison therefore illustrates both the sensitivity of $\alpha_D$ to low-energy strength and the compromise between reproducing an isolated structure and learning a shared response across nuclei. 

For applications that are particularly sensitive to this narrow structure, a combined treatment of experimental data and the ANN prediction may therefore be more appropriate. Such an approach was used in Ref.~\citep{Eizenberg:2026grw}, where we applied the ANN technique to muonic atoms. 
With the goal of providing the best possible value of $\alpha_D$, we adopt the following hybrid approach. We describe the $^{9}\mathrm{Be}$ strength below $5~\mathrm{MeV}$ with a dedicated fit to the experimental data, and retain the ANN prediction  at higher energies. Figure~\ref{fig:Be9} shows  the fit curve below 5 MeV with a dotted  line in the upper panel. Including the strength coming from such a fit to the running $\alpha_D$ sum rule, we obtained the blue curve in the bottom panel of Figure~\ref{fig:Be9}. The final value of $\alpha_D$ for $^9$Be is then 

\begin{equation}
    \alpha_D=0.548^{+0.009}_{-0.013} \, \, \mathrm{fm}^3,
\end{equation}

\noindent which is $21 \%$ larger than the value predicted by the ANN alone. This has to be considered as a realistic new update of the electric dipole polarizability for  $^9$Be.

\section{Conclusions}
\label{conclusions}
We presented an optimized ANN framework for predicting electric dipole strength functions in light nuclei with $A < 50$. The work continues the global study of Ref.~\citep{OurPaper}, but shifts the focus from coverage of the full nuclide chart to improved precision in the light nuclei region. The optimization introduces several changes that are important for this mass region. The proton number was embedded in a learned three dimensional representation, threshold information was included explicitly, the weighting was adapted to different energy regions, and the high energy response was regularized through a physics motivated loss term and a two-stage training process. The resulting ensemble yields substantially more stable strength functions and narrower ensemble intervals than the ANN of Ref.~\citep{OurPaper}. Tests on selected isotopes withheld from training show that the network reproduces the main response structures of previously unseen nuclei, supporting isotope-level transfer within elements represented in the training set.

As a benchmark application, we used the predicted strength functions to calculate the electric dipole polarizability $\alpha_D$ and compare against experimental or empirical data. We argue that the integration interval is crucial in such a comparison. When the ANN and the evaluated database entries are integrated over identical intervals, the relative  differences are small, as expected. 
However, for a correct evaluation of $\alpha_D$, the whole available energy range should be used. This is possible thanks to the continuous function provided by the ANN. We therefore provide an update on realistic $\alpha_D$ values for selected light nuclei.
In particular, for $^{9}\mathrm{Be}$, the ANN does not fully reproduce the narrow resonant structure. However, a hybrid approach that combines a dedicated fit to this structure with the smooth, continuous ANN response provides an updated estimate of the full electric dipole polarizability of 
$^9$Be.

Overall, our results support the use of the ANN as a continuous representation of the dipole response for sum-rule calculations, provided that the relevant integration limits and experimental coverage are stated explicitly. The reported HDIs quantify the variability among accepted networks under fixed choices for the data treatment, architecture, hyperparameters, loss function, and model-selection criteria; they should therefore not be interpreted as a complete propagation of experimental and modeling uncertainties. Further improvements will rely both on a more systematic treatment of experimental tensions and correlated uncertainties and, crucially, on the availability of additional data, particularly at low excitation energies. This need will become increasingly important as the framework is extended toward neutron-rich nuclei, where experimental constraints are currently sparse. New measurements for stable nuclei and unstable nuclei will provide valuable opportunities to expand the available database, test the learned isotopic trends farther from stability, and progressively improve the predictive reach of the approach.

\begin{acknowledgments}
We thank Nir Barnea,  Simone Li Muli, Immo C. Reis, and Peter von Neumann Cosel for useful discussions. 
We acknowledge support by the Deutsche Forschungsgemeinschaft (DFG) through the Cluster of Excellence ``Precision Physics, Fundamental Interactions, and Structure of Matter'' PRISMA${}^+$ EXC 2118/1 (Project ID 390831469) and through CRC1660: Hadrons and Nuclei as discovery tools (Project No. 514321794).
\end{acknowledgments}

\bibliography{bib}
\clearpage
\onecolumngrid
\appendix

\usetikzlibrary{positioning,arrows.meta,calc}

\title{Supplemental Material for ``Optimizing artificial neural networks for dipole strength predictions in light nuclei''}

\author{T.~Egert}
\email{tiegert@students.uni-mainz.de}
\affiliation{Institut f\"ur Kernphysik and PRISMA$^+$ Cluster of Excellence, Johannes Gutenberg-Universit\"at Mainz, 55128 Mainz, Germany}

\author{W.~G.~Jiang}
\email{wjiang@uni-mainz.de}
\affiliation{Institut f\"ur Kernphysik and PRISMA$^+$ Cluster of Excellence, Johannes Gutenberg-Universit\"at Mainz, 55128 Mainz, Germany}

\author{S.~Bacca}
\email{s.bacca@uni-mainz.de}
\affiliation{Institut f\"ur Kernphysik and PRISMA$^+$ Cluster of Excellence, Johannes Gutenberg-Universit\"at Mainz, 55128 Mainz, Germany}
\affiliation{Helmholtz-Institut Mainz, Johannes Gutenberg-Universit\"at Mainz, 55099 Mainz, Germany}

\maketitle

\renewcommand{\thesection}{S\arabic{section}}
\setcounter{figure}{0}
\renewcommand{\thefigure}{S\arabic{figure}}
\renewcommand{\theHfigure}{S\arabic{figure}}
\renewcommand{\thetable}{S\arabic{table}}
\renewcommand{\theequation}{S\arabic{equation}}

\section{Algorithmic and methodological details}
This Supplemental Material provides the technical details of the ANN framework summarized in the main text. It describes the learned representation of the proton number, the treatment of experimental uncertainties, the construction of the two-stage loss function, and the training and model-acceptance procedure. 

\subsection{Embedding} \label{sec::Embedding}
\begin{figure}[h]
  \centering
  \includegraphics[width=0.7\linewidth]{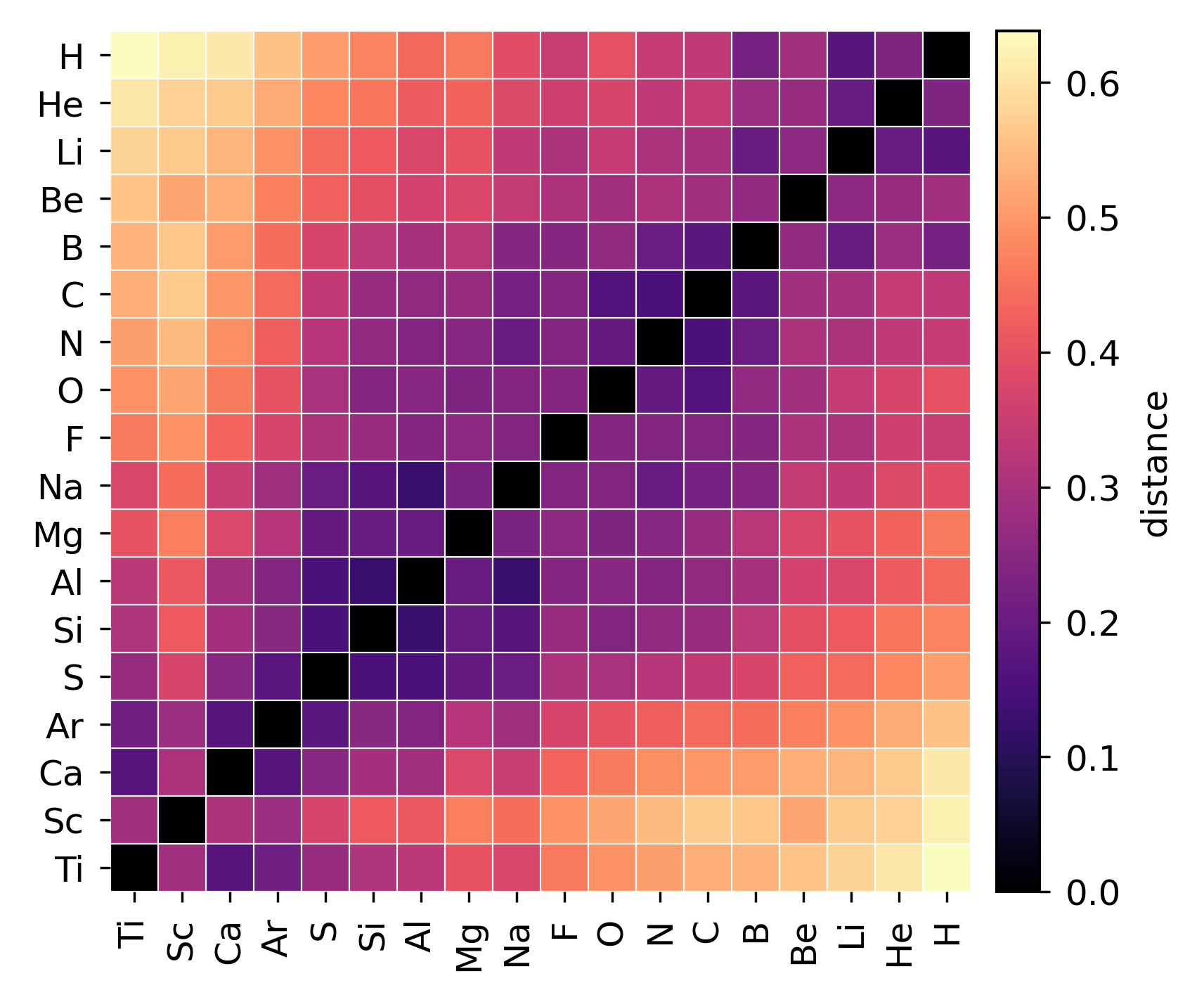}
  \caption{Mean pairwise distances between learned $Z$ embedding vectors, averaged over 100 independently trained ANN models. See text for details.}
  \label{fig::embedding}
\end{figure}

In our tests, the best trade-off was obtained by embedding only the proton number $Z$ in three dimensions. Lower dimensions yielded little benefit compared with no embedding, whereas higher dimensions degraded generalization performance. We observed a similar trend when embedding more than $Z$ (e.g., embedding both $Z$ and $A$). A plausible reason is that adding extra embedding capacity increases model flexibility and can promote overfitting to nucleus-specific patterns in the training set, thereby reducing transfer to unseen nuclei. Figure~\ref{fig::embedding} shows the mean pairwise distances between the learned $Z$ embedding vectors, averaged over the ensemble of 100 independently trained ANN models. Darker entries correspond to pairs of elements that the network represents similarly, while brighter entries indicate larger separations in the learned latent space. The matrix exhibits a pronounced banded structure, with comparatively small distances concentrated around the diagonal and generally increasing distances between elements that are further apart in nuclear charge. The banded structure is consistent with neighboring proton numbers acquiring more similar embedding vectors. However, the embedding distances alone do not determine which properties of the dipole response are encoded. The Sc nucleus forms the clearest exception to the otherwise smooth structure and remains comparatively distant even from the neighboring elements Ca and Ti. This separation may be related to the limited experimental information available for Sc. In the present dataset, Sc is constrained only by low energy Oslo method measurements, which do not directly constrain the higher energy resonance region. Its learned representation is therefore more weakly determined than those of elements for which measurements cover a broader energy range.

\subsection{Experimental weighting} \label{sec::ExpWeighting}
To account for data quality during training, the experimental uncertainties are incorporated through inverse variance sample weights. The unnormalized weight of data point $i$ is

\begin{equation}
\label{eq:errorbar}
 w_i^{\mathrm{raw}}=\frac{1}{(\delta S_{D_1,i})^2+(\delta_{i, \mathrm{min}})^2},
\end{equation}

\noindent where $\delta S_{D_1,i}$ is the experimental uncertainty of point $i$. Given that several of the older measurements do not report or estimate uncertainties, we assign an additional uncertainty of 5$\%$ of the strength function through $\delta_{i, \min}$. Without this contribution, representing an unreported uncertainty by zero would produce a formally divergent inverse-variance weight. Even after clipping, the affected data points would receive the maximum allowed weight and could dominate the training objective, leaving the network with little sensitivity to the remaining datasets. The additional term therefore permits measurements without reported uncertainties to be retained without allowing them to overwhelm measurements with quantified uncertainties. It also acts as an uncertainty floor, preventing measurements with very small quoted uncertainties from receiving excessive statistical weight. The raw weights are clipped to the interval $[10^{-1},10^{2}]$ and subsequently normalized to unit mean

\begin{equation}
 w_i=\frac{\operatorname{clip}\left(w_i^{\mathrm{raw}},10^{-1},10^{2}\right)}{\left\langle\operatorname{clip}\left(w^{\mathrm{raw}},10^{-1},10^{2}\right)\right\rangle}.
\end{equation}

\noindent The interval $[10^{-1},10^{2}]$ was chosen as a compromise between retaining the uncertainty hierarchy of the experimental data and maintaining stable optimization.

\subsection{Custom loss} \label{sec::CustomLoss}
We use two different custom loss functions, one for each training stage. The first training stage optimizes the ANN exclusively against the supervised dataset. We construct this loss $\mathcal{L}$ from three terms, $\mathcal{L}_{\mathrm{base}}$, $\mathcal{L}_{\mathrm{peak}}$, and $\mathcal{L}_{\mathrm{subthreshold}}$, which respectively fit the weighted supervised data, penalize underestimated peak strength, and suppress nonzero strength below threshold. Let $S_i$ denote the dipole strength and let $\hat S_i  =\hat S_\theta(A_i,Z_i,\omega_i)$ denote the corresponding ANN prediction. The supervised loss is then defined as
\begin{align}
\mathcal{L}^{(1)}&=\mathcal{L}_{\mathrm{base}}+ \mathcal{L}_{\mathrm{peak}} + \mathcal{L}_{\mathrm{subthreshold}} \notag\\
    &=\frac{1}{N_{\mathrm{sup}}} \sum_{i=1}^{N_{\mathrm{sup}}} w_i
    \Bigg[
    \left(S_i-\hat S_i\right)^2+ \lambda_{\mathrm{p}}\,p_i \left[S_i-\hat S_i\right]_{+}^{2}+ \lambda_{\mathrm{th}}\,s_i \left[\hat S_i-S_i\right]_{+}^{2}
    \Bigg]\,, \label{eq::stage_one}
\end{align}
where $[x]_{+}=\max(x,0)$ ensures that the peak and subthreshold contributions are one sided and become active only when the corresponding constraint is violated. $N_{\mathrm{sup}}$ is the number of supervised points and $\lambda_{\mathrm{p}}, \lambda_{\mathrm{th}}$ control the strength of the corresponding loss term. The reader should note that the displayed expression in Eq.~\eqref{eq::stage_one} gives the data-dependent loss and that the $L_2$-regularizer kernel contribution is added during optimization. The factors $p_i$ and $s_i$ are nonzero only for the selected peak and synthetic subthreshold points, respectively, and balance the influence of these points during training. To identify the peak region, a robust reference height $F_n$ is determined separately for each nucleus $n$ as the $99$th percentile of its positive target strengths between $5$ and $45~\mathrm{MeV}$. Points are classified as peak points when $8\leq\omega_i\leq40~\mathrm{MeV}$ and $S_i\geq0.9F_n$. For these points, we take

\begin{equation}
    p_i=
    \frac{N_{\mathrm{sup}}}
    {N_{\mathrm{p,nuc}}N_{\mathrm{p},n}}
    \frac{1}{w_iH_n^2}
    \left(\frac{\rho_i-0.9}{0.1}\right)^2.
\end{equation}

\noindent In this expression, $\rho_i$ is the strength relative to $F_n$, while $H_n$ is the reference height. Here, $N_{\mathrm{p,nuc}}$ denotes the number of nuclei containing selected peak points and $N_{\mathrm{p},n}$ the number of selected peak points for nucleus $n$. The normalization balances their influence nucleus by nucleus, while the final factor increases quadratically toward the resonance maximum. We set $p_i=0$ for all other points. The procedure does not identify or rank local maxima separately. Instead, it selects every supervised point within the specified energy interval whose strength exceeds $90\%$ of the robust reference height for that nucleus. For double or multi-peak responses, points associated with a secondary peak receive the auxiliary peak penalty only if they satisfy the same relative-strength criterion. Lower secondary peaks remain part of the ordinary uncertainty weighted loss but do not receive the additional one-sided peak penalty. 

The subthreshold coefficient is constructed analogously. A point is selected when it belongs to a nucleus with a known low-energy onset, satisfies $\omega_i\leq \omega_{\mathrm{th}}$, and has the synthetic target $S_i=0$. For such a point, we take

\begin{equation}
    s_i=
    \frac{N_{\mathrm{sup}}}
    {N_{\mathrm{th,nuc}}N_{\mathrm{th},n}}
    \frac{1}{w_iH_n^2},
\end{equation}

\noindent and $s_i=0$ otherwise. Here, $N_{\mathrm{th,nuc}}$ is the number of nuclei containing subthreshold data and $N_{\mathrm{th},n}$ is the number of subthreshold points for nucleus $n$. The factors proportional to $N_{\mathrm{sup}}/N_{\mathrm{p},n}$ and $N_{\mathrm{sup}}/N_{\mathrm{th},n}$ assign comparable regional loss budgets to the eligible nuclei. The factors $1/w_i$ remove the ordinary sample weighting from the effective auxiliary contributions, while $H_n^{-2}$ normalizes them relative to the characteristic response scale of each nucleus.

After successful completion of the first training stage and the associated quality checks, the model proceeds to the second stage, in which a second modified loss function is applied. Rather than imposing a common onset energy, a nucleus-dependent onset is obtained from the first-stage response and smoothed across the training nuclei. Geometrically spaced collocation pairs $(\omega_j,\omega_j^{+})$ are then generated above the measured positive-strength region and the beginning of the transition region. These points carry no prescribed strength values and constrain only the local energy dependence of the ANN response. Let $\hat S_j$ and $\hat S_j^{+}$ denote the ANN predictions at $\omega_j$ and $\omega_j^{+}$, respectively. We define

\begin{align}
    r_j &= \Delta_j+
    \frac{5}{2}\log\!\left(\frac{\omega_j^{+}}{\omega_j}\right),
\end{align}

\noindent where $\Delta_j = \log \hat S_j^{+}-\log \hat S_j,$. The residual $r_j$ vanishes when the local response follows the required $\omega^{-5/2}$ dependence, while $\Delta_j>0$ indicates a local increase in the predicted tail. The second-stage objective is

\begin{align}
    \mathcal{L}^{(2)}
    &= \mathcal{L}^{(1)}
    + \mathcal{L}_{\mathrm{pres}}
    + \mathcal{L}_{\mathrm{power}}
    + \mathcal{L}_{\mathrm{mono}}
    \notag\\
    &= \mathcal{L}^{(1)}
    + \frac{\lambda_{\mathrm{pres}}}{N_{\mathrm{sup}}}
    \sum_{i=1}^{N_{\mathrm{sup}}}
    w_i
    \left(
    \hat S_{\theta,i}-\hat S_{\theta_1,i}
    \right)^2+
    \frac{0.25}{N_{\mathrm{coll}}}
    \sum_{j=1}^{N_{\mathrm{coll}}}
    q_j
    \left[
    r_j^2+
    2\left[\max(\Delta_j,0)\right]^2
    \right]\,,
\end{align}

\noindent where $\mathcal{L}^{(1)}$ retains the supervised data term from stage one together with the peak and subthreshold penalties and $N_{\mathrm{coll}}$ is the number of tail collocation pairs. The prediction $\hat S_{\theta_1}$ denotes the accepted result of the first training stage, and the corresponding preservation term limits changes to the response in the data-constrained region. $\lambda_{\mathrm{pres}}$ controls the strength of the preservation term. The factor $q_j$ describes the smooth activation of the quasi-deuteron constraint across the transition region. For a collocation point belonging to nucleus $n$, we define

\begin{equation}
    x_j= \frac{\omega_j-\omega_{\mathrm{tr},n}}{\omega_{\mathrm{on},n}-\omega_{\mathrm{tr},n}}
\end{equation}

\noindent where $\omega_{\mathrm{tr},n}$ and $\omega_{\mathrm{on},n}$ denote the beginning and end of the transition region. The transition begins after $75\%$ of the fitted spacing between the response peak and the tail onset

\begin{equation}
    \omega_{\mathrm{tr},n}=\omega_{\mathrm{peak},n}+0.75\left(\omega_{\mathrm{on},n}-\omega_{\mathrm{peak},n} \right).
\end{equation}

\noindent The activation factor is then calculated using the smoothstep function

\begin{equation}
    q_j=x_j^2\left(3-2x_j\right).
\end{equation}

\noindent Consequently, $q_j=0$ below the transition region, increases continuously from zero to one within the transition region, and remains equal to one above the fitted onset energy. The collocation points carry a combined sample weight corresponding to $25\%$ of the supervised weight budget.

\subsection{Training}\label{sec::TrainingAndValidation}
Each ensemble member is trained in two consecutive stages. Before training, the supervised data are shuffled, and $10\%$ of the data are reserved as a fixed validation subset. During the first stage, the network parameters are optimized using the measured response data together with the synthetic zero strength points below threshold. A realization proceeds to the second stage only if the variance of the epochwise training loss is larger than $10^{-6}$. This condition rejects optimization runs for which the loss remains nearly constant. Additional nucleus by nucleus checks are applied to the threshold and peak regions. For nuclei with a known threshold, the average prediction at the synthetic zero strength points below the tabulated onset is normalized to the characteristic peak scale $H_n$. This average must remain below $1\%$ of $H_n$ for at least $90\%$ of the threshold constrained nuclei and below $5\%$ of $H_n$ for every such nucleus. Peak preservation is checked using the selected peak region points defined above. For at least $90\%$ of the nuclei, the largest prediction at these points must reach at least $65\%$ of the corresponding reference peak scale. The numerical cutoffs were introduced gradually during development, starting from loose requirements and tightening them until unwanted behavior, such as nonzero predictions below threshold, was sufficiently suppressed. The values quoted here are the final acceptance criteria used for the ensemble.

During the second stage, the same ANN is fine tuned using the supervised data together with the high energy collocation pairs. The batch normalization layers are held fixed, including their affine parameters and running statistics, while all remaining parameters continue to be optimized. After fine tuning, the nucleus wise threshold and peak checks are repeated using the final prediction. 
The high energy response is then evaluated for every training nucleus. All predicted values must remain finite. The predicted high-energy response is required to decrease smoothly and to remain consistent with the expected $\omega^{-5/2}$ QD behavior. These properties are verified explicitly for each trained model. Realizations that fail any acceptance criterion in either stage are rejected and replaced by a newly initialized training attempt. About one third of the independently trained models were accepted. The 100 accepted models all passed the threshold, peak, and high energy tail criteria described above. The most frequent stage-one failures involved excessive subthreshold strength and peak underprediction. Most stage-two rejections were caused by local increases in the high-energy tails.

\end{document}